\documentclass[aps,pra,twocolumn,showpacs,floatfix]{revtex4-1}
\usepackage[T1]{fontenc}
\usepackage{graphicx}
\usepackage{color}
\usepackage{amsmath}
\usepackage{ulem}
\usepackage{amssymb}
\usepackage{ulem}
\usepackage[usenames,dvipsnames]{xcolor}

\newlength{\halfcolumnwidth}
\begin{document}
\title{Momentum distribution of 1D Bose gases at the quasi-condensation
crossover: theoretical and experimental investigation}

\date{\today}
\author{Thibaut Jacqmin$^{(1)}$, Bess Fang$^{(1)}$, Tarik Berrada$^{(1, 2)}$, Tommaso Roscilde$^{(3)}$ and Isabelle Bouchoule$^{(1)}$}
\affiliation{$^{(1)}$Laboratoire Charles Fabry, Institut
d'Optique, CNRS UMR8501, Universit\'e Paris XI, Avenue Augustin Fresnel, 91127 Palaiseau Cedex, France\\
$^{(2)}$Vienna Center for Quantum Science and Technology, Atominstitut, TU Wien, 1020 Vienna, Austria\\
$^{(3)}$Laboratoire de Physique, CNRS UMR 5672, Ecole Normale Sup\'erieure de Lyon, Universit\'e de Lyon, 46 All\'ee d'Italie, Lyon, F-69364, France}

\begin{abstract}
We investigate the momentum distribution of weakly interacting 1D Bose
gases at thermal equilibrium both experimentally and theoretically.
Momentum distribution of single 1D Bose gases is measured using a
focusing technique, whose resolution we improve via a guiding
scheme.  The momentum distribution compares very well with quantum
Monte Carlo calculations for the Lieb-Liniger model at finite
temperature, allowing for an accurate thermometry of the gas that
agrees with (and improves upon) the thermometry based on {\it in situ}
density fluctuation measurements.  The quasi-condensation crossover is
investigated via two different experimental parameter sets,
corresponding to the two different sides of the crossover.  Classical
field theory is expected to correctly describe the quasi-condensation
crossover of weakly interacting gases.  We derive the condition of
validity of the classical field theory, and find that, in typical
experiments, interactions are too strong for this theory to be
accurate. This is confirmed by a comparison between the classical
field predictions and the numerically exact quantum Monte Carlo
calculations.

\end{abstract}

\pacs{03.75.Hh, 67.10.Ba}

\maketitle

\section{Introduction}
\label{sec.intro}

Correlation functions are essential to describe many-body systems.  In
particular, the first-order correlation function, or equivalently its Fourier
transform, the momentum distribution, is an important observable, since it
witnesses various phenomena such as the Bose-Einstein condensation transition,
the Berezinskii-Kosterlitz-Thouless transition \cite{Hadzibabic2006} or the
Mott transition \cite{Greiner2002}.  In the case of one-dimensional (1D)
homogeneous gases, although one does not expect phase transitions, the
correlation functions contain valuable information about the system.  While
the equation of state of 1D homogeneous Bose gases has been extensively
studied in various regimes~\cite{Jacqmin, amerongen08, Armijo1, Armijo2} and
compared with exact theories \cite{YangYang69}, the momentum distribution has
remained largely unexplored.  It was measured for quasi-1D quasi-condensates
using Bragg spectroscopy~\cite{Gerbier03,Hugbart2005, Richard2003}, and more
recently it was investigated using the focusing technique for quasi-1D Bose
gases in the crossover from the ideal Bose gas to quasi-condensate
regimes~\cite{amerongen08, Davis2012}.  Density ripples that appear at the
near field of a freely expanding quasi-1D quasi-condensate have also been
studied~\cite{Dettmer2001,Manz2010}.  Although related to the second-order
correlation function, they also provide information about the first-order
correlation function within the quasi-condensate theory.  The momentum
distribution of an array of strongly correlated 1D Bose gases \cite{Clement}
constitutes the only measurement on a truly 1D system to our knowledge.  From
a theoretical point of view, the momentum distribution of 1D Bose gases with
repulsive contact interactions is not known exactly over the entire phase
diagram, but a few results have been established: the mean kinetic energy can
be extracted from the exact Yang-Yang thermodynamics~\cite{Davis2012}, and the
short-range correlations are responsible for a momentum tail scaling as
$1/p^4$~\cite{Olshanii2003,Minguzzi2002,Mora03}.  The momentum distribution is
known in the asymptotic regimes such as the ideal Bose gas and the
quasi-condensate regimes, and a classical field approximation has been
proposed to account for the crossover between
them~\cite{Davis2012,Castin,Castin2,Proukakis2012}.

In this work we investigate the momentum distribution of weakly
interacting, purely or almost purely 1D Bose gases in various regimes
around the quasi-condensation crossover.  On the theory side, we find
a criterion for the classical field theory 
\cite{Castin, Castin2} to be accurate at the quasi-condensation
crossover.  Interactions are however too strong in our experiment to
fulfill this criterion, and we implement exact Quantum Monte Carlo
(QMC) simulations for the equilibrium behavior of the Lieb-Liniger
model.  We find that the measured momentum distribution agrees very
well with QMC calculations. Moreover, the temperature extracted from a
fit of the momentum distribution to the QMC calculations is in
agreement with that obtained from \textit{in situ} density fluctuation
measurements~\cite{Armijo2, Jacqmin}.

This paper is organised as follows. In Sec.~\ref{sec.theory} we recall the
classical field approximation and investigate the conditions under which it is
valid to describe the quasi-condensation crossover.  We show that, in most
experimental situations, the gas is too strongly interacting for this
description to be accurate. We then present our QMC calculations, and compare
them with the classical field prediction in the asymptotic limit.  In
Sec.~\ref{sec.technique}, we describe the experimental setup and the focusing
technique we use to measure momentum distribution. We discuss how we improve
the resolution of this technique by means of a guiding scheme.  In
Sec.~\ref{sec.results}, we present our experimental results that span both the
quasi-condensate regime (purely 1D) and the degenerate ideal Bose gas regime
(almost 1D), and compare them to QMC results.  We conclude and discuss the
prospective work in Sec.~\ref{sec.conclusions}.

\section{Theoretical predictions}
\label{sec.theory}

\subsection{Model Hamiltonian and exact solution}
\label{sec.Hamiltonian}

\begin{figure}
\includegraphics[scale=0.9]{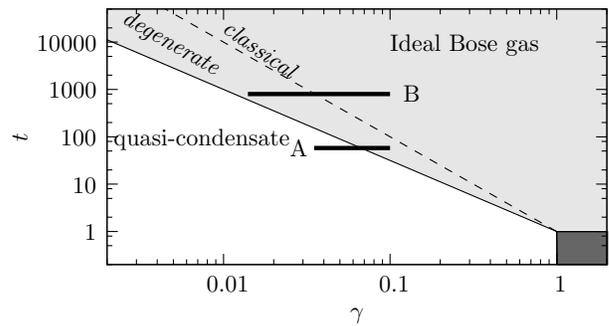}
\caption{Phase diagram of the 1D Bose gas in the parameter space
  $(\gamma, t)$.  The ideal Bose gas to quasi-condensation crossover
  occurs for $\gamma\simeq \gamma_{co}=t^{-2/3}$ (solid line).  Within
  the ideal Bose gas regime (light gray area), the crossover from the
  classical to the degenerate sub-regime occurs for $\gamma\simeq
  \gamma_d=t^{-1/2}$ (dashed line).  The region $\gamma,t \gg 1$ (dark
  gray area) is the fermionised regime.  The segments correspond to
  the regions explored by the two sets of experimental data (see
  Sec.~\ref{sec.results}).  Note that the effect of quantum
  fluctuations on the $g^{(1)}$ function in the quasi-condensate
  regime are noticeable only for extremely low temperature~\cite{TQ},
  which is experimentally irrelevant.  }
\label{fig.validity_c_field}
\end{figure}

We consider a 1D homogeneous Bose gas with repulsive contact
interactions, described by the Lieb-Liniger Hamiltonian, whose
grand-canonical expression is
\begin{equation} \label{eqn.ham}
\hat{H} =\int dz 
\left [ -\frac{\hbar^2}{2m}\hat{\Psi}^{\dagger}\frac{\partial^2}{\partial z^2}\hat{\Psi} 
+ \frac{g}{2}\hat{\Psi}^{\dagger}\hat{\Psi}^{\dagger}\hat{\Psi}\hat{\Psi} -\mu \hat{\Psi}^{\dagger}\hat{\Psi} \right ],
\end{equation}
where $z$ is the position, $\hat{\Psi}\equiv \hat{\Psi}(z)$ is the field
operator, $g$ is the coupling constant, $m$ is the mass of one boson and $\mu$
is the chemical potential. The thermal equilibrium state of the system is
completely characterised by the density $\rho$ and the temperature $T$, or by
the dimensionless interaction parameter, $\gamma = \frac{mg}{\hbar^2\rho}$,
and the reduced temperature, $t = \frac{2\hbar^2k_BT}{mg^2}$. The phase
diagram \cite{Jacqmin} in the parameter space ($\gamma$, $t$) is given in
Fig.~\ref{fig.validity_c_field}, where lines should be understood as
crossovers.  In this article, we focus on the quasi-condensation crossover,
which occurs, for $\gamma \ll 1$, around the line $\gamma_{co}=t^{-2/3}$
(solid line). The Lieb-Liniger model has the remarkable property of being
integrable, giving access to exact results valid over the entire phase
diagram. For example, the equation of state is known exactly through the
Yang-Yang theory \cite{YangYang69}. Yet, the calculation of the first order
correlation function $g^{(1)}(z) = {\langle \hat{\Psi}^\dagger
  (z)\hat{\Psi}(0)\rangle}$ from the Bethe-Ansatz solution for finite $\gamma$
and at finite temperature is still a subject of active research.  Approximate
analytic theories for the correlation functions exist in the limiting cases of
ideal Bose gas regime (where the ideal Bose gas theory applies) and
quasi-condensate regime (where the Bogoliubov theory applies \cite{Mora03}).
However, the behavior of $g^{(1)}(z)$ at the crossover between those two
regimes lacks a quantitative description.

\subsection{Classical field theory}
\label{sec.CFT}

The full quantum many-body problem of Eq.~(\ref{eqn.ham}) is notoriously
complex.  However, since thermal fluctuations are expected to dominate at the
quasi-condensation crossover, it seems appropriate to simplify the problem
using a classical field approximation, where the quantum field operators
$\hat{\Psi}(z)$ and $\hat{\Psi}^\dagger (z)$ are replaced by the complex
fields $\Psi(z)$ and $\Psi^*(z)$ \cite{Castin, Castin2}. This strategy has
been pursued in \cite{Proukakis,Davis2012,Proukakis2012}.  Such an approach is
particularly attractive for 1D homogeneous gases: contrary to higher
dimensions, no ultraviolet divergence occurs when calculating $g^{(1)}(z)$, so
that no energy cut-off is required for this particular quantity.  This problem
can then be mapped onto a time-independent 2D Schr\"odinger problem,
describing a single particle evolving in imaginary time~\cite{Castin}, thus
leading to fast and easy-to-implement calculations.

The classical field model is parametrised by a single dimensionless variable
$\nu=\gamma/\gamma_{co}=\gamma t^{2/3}$, provided that lengths are rescaled by
the correlation length $l_c=\hbar^2\rho/(mk_B T)$~\cite{Castin, correlk}.
Simple analytic formulas are found in the two limits of the ideal Bose gas
regime ($\nu \gg 1$) and the quasi-condensate regime ($\nu \ll 1$): the
momentum distribution is Lorentzian in both limits, with its full width at
half maximum (FWHM) being $\Delta p = 2\hbar/l_c$ in the ideal gas limit and
$\Delta p=\hbar/l_c$ in the quasi-condensate limit.  One recovers here the
expressions derived from the highly degenerate ideal Bose gas model and the
high-temperature Bogoliubov theory~\cite{Mora03} respectively.  Results of
numerical calculations spanning the crossover between those two regimes are
reported in Fig.~\ref{fig.compQMC}, where one assumes a fixed value $t=1000$
and parametrises the system with $\gamma$ instead of $\nu$.  We define the
crossover region as the domain for which $\Delta p$ (shown in
Fig.~\ref{fig.compQMC}(a) as the dashed line), differs by more than 10\% from
both asymptotic limits ($\hbar/l_c$ and 2$\hbar/l_c$), and we find that the
crossover extends over about one order of magnitude in $\nu$, or equivalently
in $\gamma$ at fixed $t$.  Looking at the shape of the momentum distribution,
we recover the Lorentzian distribution in both asymptotic regimes and, for any
value of $\nu$, we find momentum tails decreasing as $1/p^2$, as expected from
the analog single-particle problem~\cite{SP}.

\subsection{Validity of the classical field theory at the quasi-condensation
crossover}
\label{sec.CFTvalidity}

The classical field theory is valid only if the population
of the relevant modes is large.  Within the ideal Bose gas regime,
this requires $\rho\gg \rho_{d}=\sqrt{mk_BT}/\hbar$, {\it i.e.}  a
highly degenerate gas. This condition writes equivalently $\gamma\ll
\gamma_d$ where $\gamma_d=t^{-1/2}$ is the interaction parameter at
degeneracy.  Thus, the classical field theory only correctly
describes the quasi-condensation crossover (which occurs at
$\gamma\simeq\gamma_{co}$) provided that $\gamma_{co}\ll \gamma_d$.
The last condition translates into $t^{1/6}\gg 1$. Since crossovers
span typically about one order of magnitude in $\gamma$, one
requires that $t\gtrsim 10^6$.  This value is very difficult to
achieve experimentally on cold-atoms experiments while
maintaining a temperature sufficiently low to ensure the 1D
condition, $k_B T \ll \hbar \omega_{\perp}$, unless
extremely weak atomic interactions are reached, using for
instance a Feshbach resonance~\cite{Bouchoule2007}.

\begin{figure*}
\centerline{\raisebox{-0.08cm}{\includegraphics[scale = 0.8]{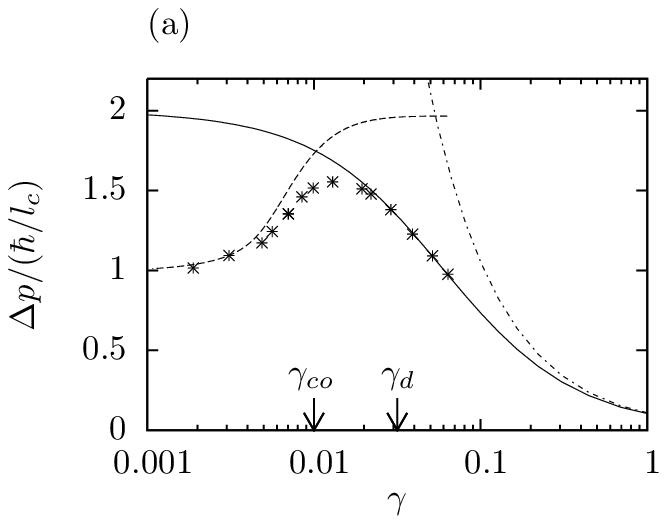}}
\includegraphics[scale = 0.8]{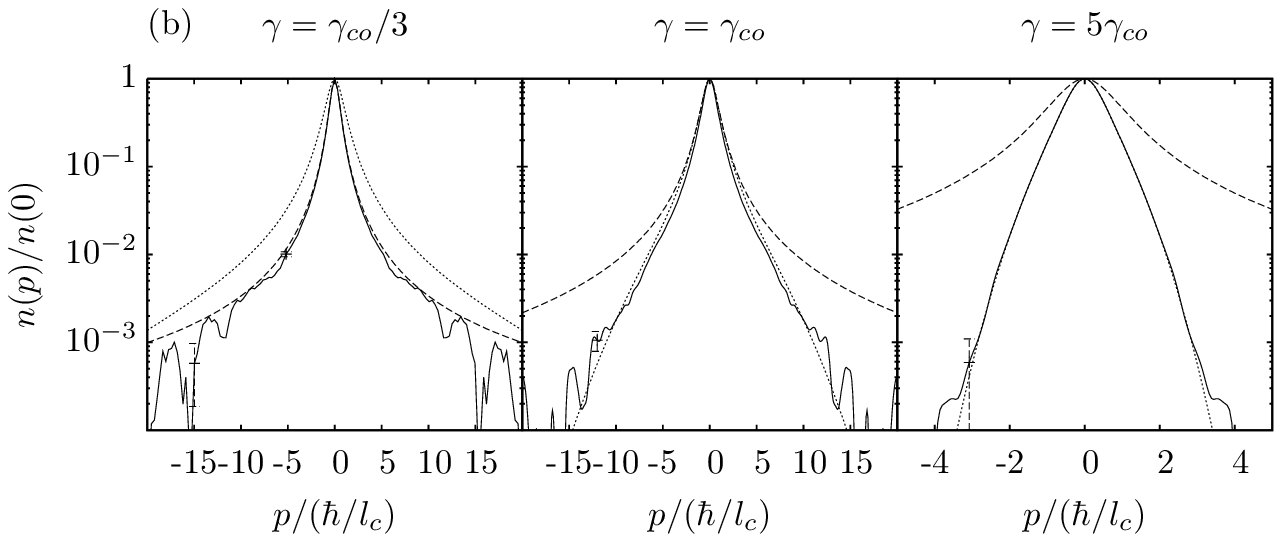}}
\caption{Momentum distribution of a homogeneous gas at a $t=1000$.
  (a) FWHM according to the classical field theory (dashed line),
  ideal Bose gas theory (solid line), Maxwell-Boltzmann prediction
  (dash-dotted line) and QMC calculations (stars).  The arrows show
  the interaction parameter at the quasi-condensation crossover
  ($\gamma_{co}=t^{-2/3}$) and at the degeneracy crossover
  ($\gamma_d=t^{-1/2}$).  (b) momentum distribution in log scale for
  three different value of $\gamma$. Shown are QMC results (solid
  lines), classical field predictions (dashed lines) and ideal Bose
  gas prediction (dotted lines). Error bars, shown for a single 
point on the side of the profile, corresponds to 
one standart deviation of the QMC calculation.  }
\label{fig.compQMC}
\end{figure*}

In the experiment presented here, we have $t<1000$.  In this case, according
to the above argument, the classical field approach is expected to be
inaccurate at the quasi-condensation crossover.  This can be seen in
Fig.~\ref{fig.compQMC}(a), which shows the FWHM of the momentum distribution
as a function of $\gamma$ according to the ideal Bose gas theory and the
classical field prediction for $t=1000$.  The ideal Bose gas theory is
parametrised by $\chi = \gamma/\gamma_d\simeq \rho_d/\rho$, which quantifies
the level of degeneracy: the FWHM of the momentum distribution within this
theory goes from the Maxwell-Boltzmann prediction for $\chi\gg 1$ to
$2\hbar/l_c$ for $\chi \ll 1$, and this crossover spans more than one order of
magnitude in $\chi$, or equivalently in $\gamma$ at fixed $t$.  The classical
field theory correctly describes the quasi-condensation crossover only if it
shares a common plateau with the ideal Bose gas theory at $\Delta p
=2\hbar/l_c$ in the degenerate ideal Bose gas regime, where $\gamma_{co} \ll
\gamma \ll \gamma_d$.  As seen in Fig.~\ref{fig.compQMC}(a), this is not the
case at all for $t=1000$: the highly degenerate ideal Bose gas regime is not
very well identified for such a small value of $t$ as far as the momentum
distribution is concerned, so that the classical field theory does not
correctly describe the quasi-condensation crossover.

\subsection{Quantum Monte Carlo approach}
\label{sec.QMC}

For $t<10^6$, one needs therefore a more accurate treatment than the
above-mentioned classical field approximation.  We make use of QMC
simulations to provide numerically exact results for the Lieb-Liniger
model at finite temperature.  Bosons in continuous space can be
efficiently simulated making use of the density matrix renormalization
group in one dimension \cite{Schmidt} and Monte Carlo techniques in
arbitrary dimensions.  Variational and diffusion Monte Carlo results
for the first-order correlation function at $T=0$ are reported in
Ref.~\cite{AstrakharchikG03}.  At finite temperature, one can in
principle make use of path-integral Monte Carlo simulations
\cite{Ceperley95,Boninsegnietal06}, although we are not aware of
explicit calculations for the Lieb-Liniger model. This latter
technique requires the discretization of the imaginary-time dimension
(Trotter approximation), and its influence on the simulation results
has to be carefully removed. Here, we adopt a rather complementary
approach, discretizing the spatial dimension instead
\cite{Schmidt}. Indeed, bosonic lattice models lend themselves to
efficient QMC simulations free of any Trotter approximation. By
discretizing the field operator $\hat{\psi}^{(\dagger)}(z) \approx
\hat{b}_j^{(\dagger)}/\sqrt{a}$, where $a$ is the lattice spacing,
$\hat{b}_j$, $\hat{b}_j^{\dagger}$ are bosonic operators at site $j$,
and $z \approx j a$, we obtain the Bose-Hubbard Hamiltonian
\begin{eqnarray}
\hat{H}_{\rm BH} = \sum_j &\Big[ &-J \left(\hat{b}_{j+1}^{\dagger}
  \hat{b}_j + {\rm h. c. }\right) + \frac{U}{2} \hat{n}_j (\hat{n}_j -
  1) \nonumber \\ && - (\tilde{\mu } - v_j) ~\hat{n}_j ~\Big],
\end{eqnarray}
with $\hat{n}_j = \hat{b}^\dagger_j \hat{b}_j$, and parameters $J =
\hbar^2/(2ma^2)$, $U = g/a$, $\tilde{\mu} = \mu - 2J$. We also include
the presence of an external potential $V(z)$, discretized to give
$v_j$.  The Lieb-Liniger parameters $\gamma$ and $t$ read
\begin{equation}
\gamma = \frac{U}{2J n}~, ~~{\rm and}~~~ t = \frac{4 k_B T}{J} \left(
\frac{J}{U} \right)^2,
\end{equation}
where $n = \langle \hat{n}_j \rangle = \rho a$ is the lattice filling. 

We study the Bose-Hubbard approximation to the Lieb-Liniger Hamiltonian by
making use of Stochastic Series Expansion (SSE) QMC
\cite{Syljuasen03,Sandvik10}, extensively used to investigate lattice
bosons. The Lieb-Liniger model is correctly recovered when the lattice spacing
is much smaller than the correlation length in the system. In the classical
ideal Bose gas regime, this condition amounts to require $\lambda_{\rm dB}/a =
\sqrt{4\pi J/(k_B T)} \gg 1$. On the other hand, in the degenerate ideal Bose
gas regime and quasi-condensate regime, we need $l_c / a \gg 1$, implying
$2J/(U\gamma t) \gg 1$ and $4J/(U\gamma t) \gg 1$ respectively. The
first-order correlation function and the momentum distribution are efficiently sampled
with SSE-QMC during the directed-loop update \cite{Syljuasen03,Sandvik10}. The
lattice simulation reproduces faithfully the momentum distribution of the
Lieb-Liniger model for momentum $p \ll \hbar\pi/a$.
 
Fig.~\ref{fig.compQMC} shows QMC calculations for a homogeneous gas at
$t=1000$, for values of $\gamma$ that span the quasi-condensation
crossover.  In Fig.~\ref{fig.compQMC}(a) the FWHM of the momentum
distribution calculated with QMC is compared to that of the classical
field and of the ideal Bose gas models.  The QMC results follow the
ideal Bose gas prediction almost until the FWHM of the ideal Bose gas
prediction crosses that of the classical field prediction at
$\gamma=\gamma_c$, and then converges towards the classical field
prediction.  The disagreement between QMC and the above theories never
exceeds 20\%.  These results mitigate the ``failure'' of the classical
field approximation: as long as the FWHM is concerned, an approximate
model where one uses the ideal Bose gas prediction for
$\gamma>\gamma_c$ and the classical field prediction for
$\gamma<\gamma_c$, would give the correct predictions within 20\%
error.  Investigation of the full momentum distribution is shown in
Fig.~\ref{fig.compQMC}(b) for three different values of $\gamma$.  On
the ideal Bose gas side of the quasi-condensation crossover, for
$\gamma=5\gamma_{co}$, the QMC momentum distribution follows closely
the ideal Bose gas behaviour. The classical field result is recovered
for $\gamma\approx \gamma_{co}/3$, while the momentum width is still
about 10\% above the Bogoliubov prediction. For $\gamma=\gamma_{co}$,
we find that the tails of the momentum distribution agree well with
the ideal Bose gas prediction, while the width of the distribution is
narrowed by about 20\%.  Finally, note that, although the highly
degenerate ideal Bose gas prediction $\Delta p =2\hbar/l_c$ has not
been reached before the gas undergoes the quasi-condensation
crossover, the momentum distribution at $\gamma\simeq \gamma_{co}$ is
much narrower than the Maxwell-Boltzmann prediction (dash-dotted line
in Fig.~\ref{fig.compQMC}(a)), so that the effect of degeneracy within
the ideal Bose gas regime is already substantial.

\subsection{ Local density approximation}
\label{sec.LDA}

While the previous results hold for homogeneous infinite systems, in
the experiment described below, the gas is trapped longitudinally in a
harmonic confinement of frequency $\omega/(2\pi)$.  As long as the
correlation length $l_c$ is much smaller than the cloud size $L$,
however, a local density approximation (LDA) is valid: at a given
position $z$, the local value of an observable $O$ is that of a
homogeneous gas ($O_{\rm hom}$) at the chemical potential $\mu_{\rm
  loc}(z) = \mu - m\omega^2 z^2/2$, and the global value for a trapped
system ($O_{\rm trap}$) can be obtained by adding the contributions of
each slice of the gas, \textit{i.e.},
\begin{equation}
 O_{\rm trap}(\omega,\mu) = \sqrt{\frac{2}{{m\omega^2}}} \int_0^{\infty}
 \frac{d{\tilde\mu}}{\sqrt{\tilde{\mu}}} ~O_{\rm hom}(\mu-\tilde\mu)~.
 \end{equation}
Therefore, within this approximation and for a given value of $\mu$
(fixing the peak density), the normalized momentum distribution is
independent of $\omega$, while the total atom number scales as
$1/\omega$.  For a gas whose peak linear density lies in the crossover
from the degenerate ideal Bose gas regime to the quasi-condensate
regime, the LDA condition $l_c\ll L$ writes $\omega \ll
\left(mg^2k_B^2T^2/\hbar^5\right)^{1/3}$ \cite{Bouchoule2007}.  At
much higher densities, when almost the whole cloud lies within the
quasi-condensate regime, the LDA condition gives $\omega \ll
k_BT\sqrt{2mg/\rho}/\hbar^2$ \cite{Gerbier03}.  Both conditions are
fulfilled in the data shown below, so that the LDA is expected to be
valid.  Using QMC, we compute the momentum distribution (normalised
to the total atom number) of a
harmonically trapped gas for different chemical potentials ({\it i.e.}
different peak linear densities) and temperatures. To shorten the
computational time, calculations are performed for trapping
frequencies five times larger than the experimental values, while
still satisfying the LDA condition, as verified numerically.
The computed momentum distribution
are multiplied by five to be compared to experimental data.
  Linear interpolation between the calculated distributions permits the
calculation of the momentum distribution for any chemical potential
$\mu$ and temperature $T$, and enables us to perform fits to the
measured distribution. We check {\it a posteriori}  that 
the interpolation is correct up to a few percent.

\section{Experimental techniques}
\label{sec.technique}

\begin{figure}
\includegraphics[scale=0.16]{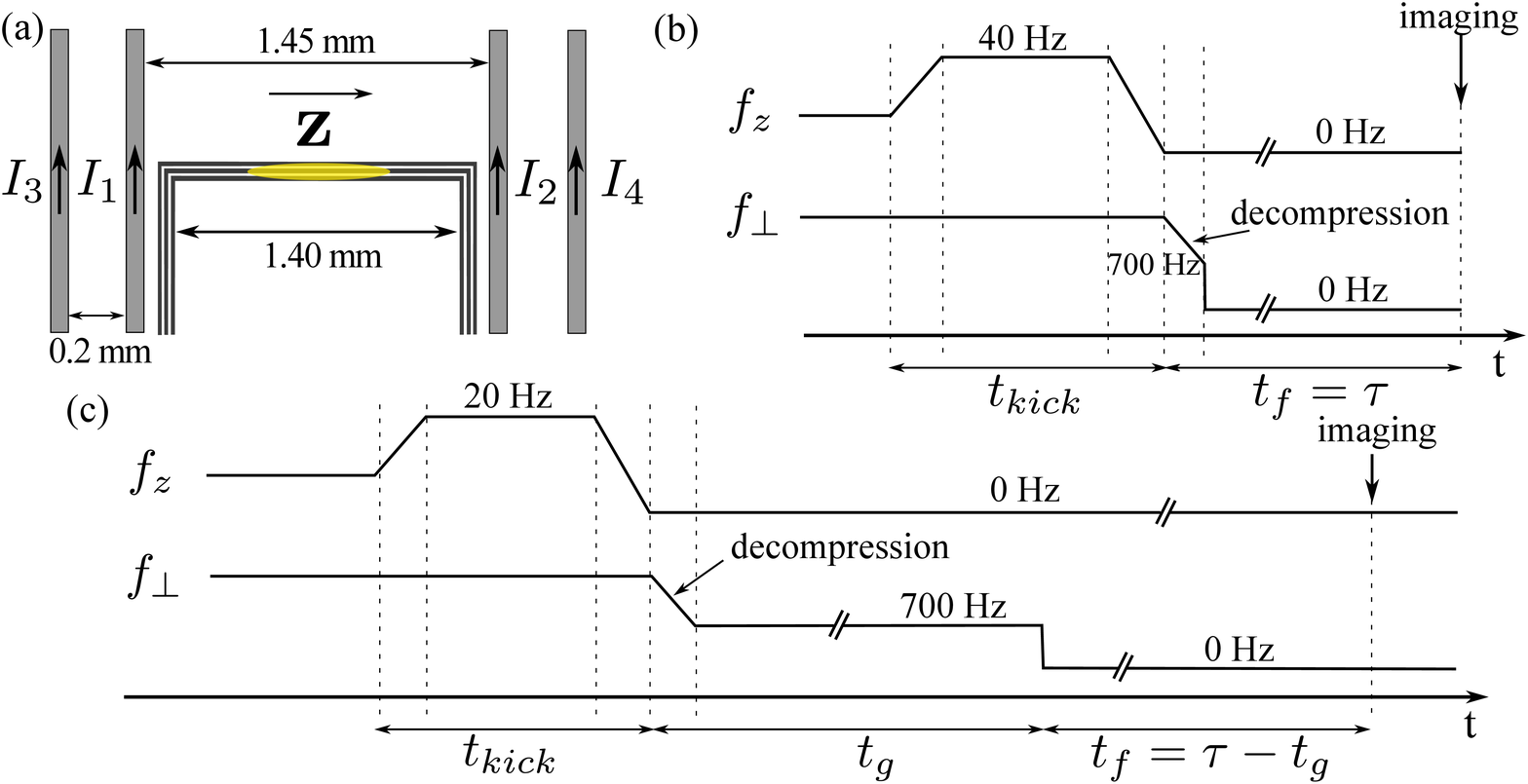} 
\caption{(a) Wire configuration: currents $I_1$, $I_2$, $I_3$ and
  $I_4$ realise the longitudinal trapping potential of frequency
  $f_z$.  The transverse confinement of frequency $f_\perp$ is ensured
  by the three wires carrying a current modulated at $200$~kHz.  (b,c)
  Focusing sequences.  The focusing potential, harmonic to order $5$
  in $z$, is applied during $t_{\rm kick}$ before the longitudinal
  potential is turned off. After a free evolution during the focusing
  time $\tau$, an absorption image is taken.  Without guiding, see
  (b), the transverse confinement is removed during the whole focusing
  time, though a 1~ms ramp down to $f_{\perp}=700$~Hz is carried out
  to decrease the transverse expansion.  In the presence of
  guiding, see (c), $f_{\perp}$ is kept at $700$~Hz for a time
  $t_g$, during which the cloud does not fall.  Numerical values in
  (b) and (c) correspond to data of Fig.~\ref{fig.results} and
  \ref{fig.levitation} respectively.  }
\label{fig.config_fils}
\end{figure}

One-dimensional atomic clouds are realised in our experiment using an
atom-chip setup \cite{Reichel}.  More specifically, we load $8 \times 10^4$
Rb$^{87}$ atoms in the modulated guide described in \cite{Jacqmin}.  The
longitudinal harmonic potential has an oscillation frequency of about $8$~Hz
and is obtained by DC currents running through wires $I_3$ and $I_4$ (see
Fig.~\ref{fig.config_fils}). The transverse potential is provided by three
wires carrying a $200$~kHz sinusoidal current.  The transverse oscillation
frequency can be varied from a few hundred ~Hz to tens of ~kHz by changing the
current amplitude.  An important feature of this design is the independence
between the longitudinal and transverse confinements.  After $1200$~ms of
radio-frequency evaporation, we let the cloud reach thermal equilibrium for
$400$~ms, after which no remaining breathing can be observed.  To characterise
the atomic cloud, we perform \textit{in situ} density fluctuations
measurements, analysing the noise in an ensemble of absorption images as
described in~\cite{Armijo1}.

To measure the momentum distribution, one can monitor the free expansion of
the gas after all confining potentials have been turned off.  The rapid
transverse expansion amounts to an effective instantaneous suppression of the
interactions, so that the longitudinal expansion reflects faithfully the
initial momentum distribution in the trap.  However, reaching the far field
regime where the longitudinal density profile is homothetic to the momentum
distribution requires unrealistically large field of view and long expansion
time for our experimental parameters.  We thus use the so called focusing
technique, already applied to 1D \cite{Shvarchuck, VanEs2010, Davis2012} and
2D systems \cite{Cornell}. Its principle is described below.  A strong
longitudinal harmonic potential is applied for a very short time during which
atoms do not have time to move but acquire a longitudinal momentum shift
$\delta p = -Az$, proportional to their distance $z$ to the trap center. Then,
the atomic interactions are suppressed by switching off the transverse
confinement, and the longitudinal confinement is removed so that the cloud
starts a free evolution.  After a focusing time $\tau = m/A$, the longitudinal
density distribution $f(z)$ is homothetic to the initial momentum distribution
$n(p)=(m/\tau)f(p\tau/m)$~\cite{ntof}.  More precisely, the time sequence we
use is drawn in Fig.~\ref{fig.config_fils}.  The longitudinal focusing
potential is realized by a four-wire configuration that cancels out the non
harmonic terms up to the 6th order, thus minimizing aberrations.  It has an
oscillation frequency $\omega_{\rm{kick}}/2\pi=40$~Hz and it is applied for
about $t_{\rm{kick}}=0.6$~ms.  After the focusing pulse, we ramp the
transverse frequency down to $700$~Hz in $1$~ms before completely turning off
the transverse trapping potentials and letting the cloud evolve freely. The
$1$~ms transverse ramp is used to limit the transverse velocity spread, so
that the transverse final expansion of the cloud is reduced and the signal to
noise ratio (SNR) is increased.  We verified that this ramp is however quick
enough to leave the longitudinal velocity distribution unchanged.  After free
evolution during the focusing time $\tau\simeq
1/\omega_{\rm{kick}}^2t_{\rm{kick}}\approx 27$~ms, we take an absorption image
of the cloud and extract the longitudinal density distribution, from which we
deduce the initial momentum distribution.  The focusing time $\tau$ is
adjusted by minimising the width of the density distribution of cold samples.
Moreover, the width of very cold clouds saturates at a value that gives us the
momentum resolution $\Delta p/ \hbar = 0.2~\mu {\rm m}^{-1}$.  This resolution
is significantly smaller than the width of the data in Fig.~\ref{fig.results}.

\begin{figure}[ht]
\centering{
\includegraphics[scale=0.5]{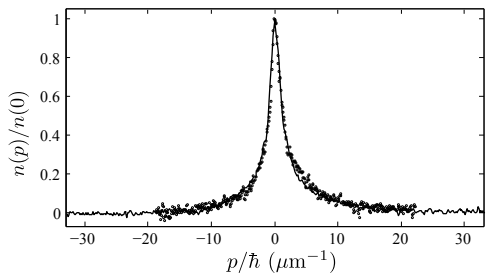}
}
\caption{Momentum distribution of a gas of about $7000$ atoms at
  thermal equilibrium, in a $7.5$~Hz longitudinal trap, and $4.6$~kHz
  transverse trap. The solid distribution is obtained without
  guiding and with a free fall of $24.7$~ms. The circles are the
  momentum distribution of the same sample, but with a guiding time
  of $20.6$~ms followed by a free fall of $22.7$~ms. }
\label{fig.levitation}
\end{figure}

On the other hand, a better resolution may be necessary for a momentum
distribution that is much narrower.  This can be achieved by a longer focusing
time $\tau$.  However, the depth of field of our optical system limits the
maximum free fall under gravity to about $3$~mm, corresponding to a free fall
time of about $25$~ms.  Nevertheless, larger focusing times can be obtained if
a transverse potential is maintained during the focusing time, strong enough
to overcome gravity (which acts in the transverse plane in our experiment),
but weak enough so that atomic interactions are negligible. More precisely the
guiding scheme we use is the following.  After having performed the
longitudinal focusing pulse and switched off the longitudinal potential, we
ramp the transverse frequency down in $1$~ms, similar to the procedure
described before.  The final transverse frequency is $f_l=700$~Hz, which is
the lowest value that allows the atoms to remain transversally trapped.  We
then hold the cloud for a guiding time $t_g$, at the end of which we switch
off the transverse confinement and let the cloud fall freely for $t_f$. The
guiding does not affect the longitudinal free evolution as long as the
effect of interactions stays negligible. We have checked that this is indeed
the case for the explored parameter range by measuring the longitudinal
density profile at focus $f(z)$ of identical clouds, either without guiding
and a focusing time $\tau=t_f =24.7$~ms, or with the above guiding scheme
and $\tau = t_g+t_f = 43.3$~ms, where $t_g = 20.6$~ms and $t_f = 22.7$~ms.
The functions $f(mz/\tau)$, shown on Fig.~\ref{fig.levitation} for a cloud of
$7000$ atoms at $T\approx 95$~nK, give indeed the same momentum distribution.
The validation of this technique paves the way towards high resolution
measurements of momentum distribution of 1D gases.

\section{Experimental results}
\label{sec.results}

\begin{figure}
\includegraphics[scale = 1]{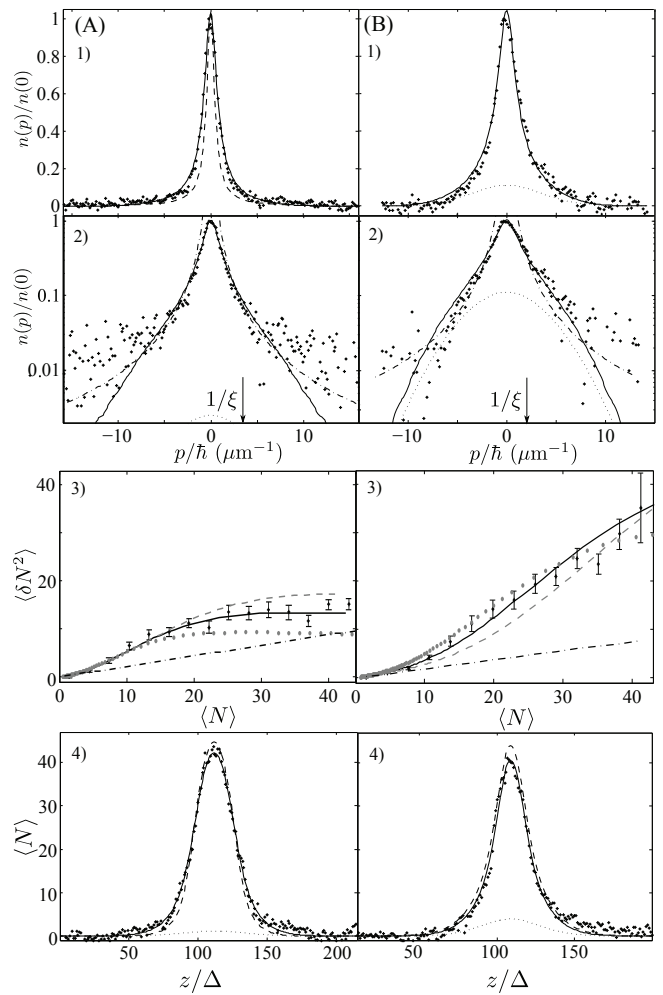} 
\caption{Data for atoms confined in a harmonic trap of transverse
  oscillation frequency of (A) 6.4~kHz and (B) 2.1~kHz, and a
  longitudinal oscillation frequency of (A) 8.3~Hz and (B) 7.6~Hz.
  The pixel size, in the object plane, 
  for \textit{in situ} data is $\Delta=2.7~\mu$m.
  1),2) momentum distribution in linear and log scales.  QMC fits are
  shown in solid lines, yielding (A) $T=72$~nK and (B) $T=84$~nK.  The
  dashed line in (A,1) is the prediction using the LDA and the
  Lorentzian momentum distribution in the quasi-condensate limit.  In
  2), the dashed-dotted lines give the $1/p^2$ behavior, while the
  momentum associated with the healing length $\xi=\hbar/\sqrt{m \rho
    g}$ indicates that sample (A) is more in the quasi-condensate,
  with $\Delta p \ll\hbar/\xi$, and sample (B) is more in the ideal
  Bose gas, with $\Delta p \gg\hbar/\xi$.  In addition, the dotted
  lines indicate the contribution from the transverse excited states,
  which is insignificant for data (A) and hence neglected in our
  analysis.  3) \textit{in situ} density fluctuations, with Yang-Yang
  prediction in solid lines, evaluated at the above-mentioned
  temperatures.  The gray dashed and dotted lines shows Yang-Yang
  predictions with 30\% temperature deviations.  The dash-dotted lines
  indicate the poissonian fluctuation level.  4) \textit{in situ}
  profile, with the QMC profile in dashed lines.  The dotted lines
  again show the contribution from transverse excited states.  We also
  plot in solid lines the density profile fitted with Yang-Yang
  calculation, which yields (A) $T=111$~nK and (B) $T=76$~nK.  }
 \label{fig.results}
\end{figure}

Fig.~\ref{fig.results} shows experimental results referring to two different
parameter sets (particle number, temperature and trapping potential).  In both
cases, the momentum distribution is obtained without the guiding scheme,
with a focusing time of $27$~ms.

\subsection{On the quasi-condensate side of the crossover}

Data in Fig.~\ref{fig.results}(A) correspond to a cloud initially trapped with
a transverse oscillation frequency $\omega_\perp/(2\pi)=6.4$~kHz and a
longitudinal frequency of $8.3$~Hz, measured using parametric heating and
dipole oscillations respectively.  Since the cloud temperature and chemical
potential are much smaller than the transverse level spacing, the system is well
described by a 1D gas with a coupling constant $g=2\hbar\omega_\perp a_s$,
where $a_s=5.3~$nm is the 3D scattering length~\cite{Olshanii1998}.  In
Fig.~\ref{fig.results} (A,1-2), we show the measured momentum distribution
(points) with a QMC fit (solid lines) that yields $T=72$~nK, corresponding to
$t=76$.  Note that the shape is reproduced up to a few per cent.  We report in
Fig.~\ref{fig.validity_c_field} the segment $[(\gamma_m,t),(\gamma_M,t)]$
corresponding to the domain explored by the data: $\gamma_m$ corresponds to
the peak density, and the segment length is such that 80\% of the atoms lie in
a region of density larger than that associated to $\gamma_M$.  We find that
$\gamma_m\simeq \gamma_{co}/2$, where $\gamma_{co}=t^{-2/3}$ is the
interaction parameter at the quasi-condensation crossover.  Thus the cloud
lies quite deeply in the quasi-condensate regime. However, the measured
momentum distribution is substantially broader than that predicted using the
quasi-condensate Lorentzian $n(p) \propto1/(p^2+(\hbar/l_c)^2)$ together with
the LDA, shown as the dashed line in Fig.~\ref{fig.results}(A,1).  On the
other hand, for such a value of $\gamma_m$, the momentum distribution shows
slowly decreasing tails, compatible with a Lorentzian behavior within a few
percent in a wide range of momenta. This is seen in
Fig.~\ref{fig.results}(A,2) where the dashed-dotted line shows a $1/p^2$
decrease.  For our SNR of about 80, data are compatible with a Lorentzian
behavior at large $p$.
 
As shown in~\cite{Armijo2,Jacqmin}, an independent and reliable thermometry
can be carried out using \textit{in situ} density fluctuations, provided the
pixel size $\Delta$ is both much larger than the correlation length of density
fluctuations, and much smaller than the typical length scale of the variation
of the mean density profile~: the atom number fluctuations are then given by
the thermodynamic relation $\langle \delta N^2\rangle= k_B T\Delta \partial
\rho/\partial \mu$, where $\rho(\mu,T)$ is the equation of state of
homogeneous 1D gases, known exactly using the Yang-Yang
calculation~\cite{Armijo2}, and the derivative is computed for $\mu$ such that
$\rho(\mu,T)=\langle N \rangle /\Delta$ is the local density.  The above
criterion is well fulfilled in our experiment and the measured density
fluctuations are shown in Fig.~\ref{fig.results}(A,3).  For linear densities
larger than 25 atoms per pixel, fluctuations are almost independent of the
density, which confirms that the center part of the gas lies within the
quasi-condensate regime. Indeed, in this regime $\mu\simeq g\rho$, so that the
thermodynamic relation reduces to $\langle \delta N^2\rangle=\Delta k_B T/g$,
which does not depend on $\langle N \rangle$.  The fluctuations computed using
the thermodynamic relation and the Yang-Yang equation of state, for the
temperature $T=72$~nK (obtained by fitting the momentum distribution with the
QMC results) are shown as solid line.  The data are in agreement with this
prediction.  With our SNR, the density fluctuations based thermometry has a
precision of about 20\% (see gray curves in Fig.~\ref{fig.results}(A,3)), less
precise than the thermometry obtained by fitting QMC calculations to the
momentum distribution.

Finally, we compare in Fig.~\ref{fig.results}(A,4) the measured \textit{in
  situ} density profile to the QMC density profile (within LDA) for $T=72$~nK
(dashed line). We find a good agreement for most of the profile, although the
measured data show higher wings.  Since the central part of the cloud lies
deep in  the quasi-condensate regime, a large part of the cloud follows the
Thomas-Fermi profile, and all the information on the temperature lies in the
\textit{small} wings. This renders Yang-Yang thermometry based on density
profile less precise and extremely sensitive to the shape of the wings. Here,
a Yang-Yang fit to the density profile gives $T=110$~nK, a value 40\% higher
than that extracted from the QMC fit, incompatible with the measured momentum
distribution or the density fluctuation measurements. This discrepancy, and
thus the presence of the inflated wings, may come from the anharmonicity of
the potential due to its residual roughness~\cite{anharm}.  Alternatively, it
may also indicate a lack of perfect thermal equilibrium.

\subsection{On the ideal Bose gas side of the crossover}
\label{sec.resultsB}

While the previous results probe mainly the quasi-condensate regime,
we also probe the ideal Bose gas side of the quasi-condensation
crossover, {\it i.e.}  data with $\gamma_m >\gamma_{co}=
t^{-2/3}$. The ideal Bose gas regime shows a very different behaviour
from the trivial Maxwell-Boltzmann prediction only for a large $t$
parameter, for which the quasi-condensation crossover occurs for an
already highly degenerate gas.  If one wishes to preserve the 1D
condition $k_B T \ll \hbar \omega_\perp$, large $t$ parameters can be
accessed only by decreasing the transverse
confinement~\cite{Bouchoule2007}.  We thus reduced the transverse
confinement to 2.1~kHz. Data are shown in Fig.~\ref{fig.results}(B).
No saturation of the density fluctuations is seen on the \textit{in
  situ} fluctuation measurements, which indicates that the gas does
not lie in the quasi-condensate regime. Fluctuations however rise well
above the poissonian level (shown as dashed-dotted line on
Fig.~\ref{fig.results}(B,3)) so that the gas is highly degenerate.
Contrary to the data of Fig.~\ref{fig.results}(A), we now have
$k_BT/(\hbar\omega_\perp)=0.8$ which is of the order of unity, so that
the transverse excited states contribute to the measured fluctuations
and momentum distribution. We take into account the population of
transverse excited states, assuming that they behave as independent
ideal 1D Bose gases, while the transverse ground state is treated 
as a Lieb-Liniger gas.
This modified Lieb-Liniger model has been used with
success to describe density profiles~\cite{amerongen08} and density
fluctuations~\cite{Armijo2}, and has been applied to predict the rms
width of momentum distribution~\cite{Davis2012}. 
Here we use our QMC calculations
to describe the transverse ground state.
A fit using the above
strategy reproduces well the measured momentum distribution (see
Fig.~\ref{fig.results}(B,1)), and yields the temperature $T = 84$~nK,
corresponding to $t = 840$. The segment in the phase space
$(\gamma,t)$ explored by the data is shown in
Fig.~\ref{fig.validity_c_field}: the peak linear density is close to
the quasi-condensation crossover density and most of the cloud lies in
the degenerate ideal Bose gas regime.  The contribution of the excited
states to the momentum distribution is seen in
Fig.~\ref{fig.results}(B,1): it is only 10\% in the center but it
rises to almost 50\% in the wings around $|p|/\hbar \simeq 5
~\mu\mbox{\rm m}^{-1}$. Fig.~\ref{fig.results}(B,2) shows that the
theoretical momentum distribution decreases faster than $1/p^2$ at
large momenta.  This is mainly due to the contribution of the excited
states, which have approximately Gaussian momentum tails. However,
with our SNR of about 50, no deviation from a $1/p^2$ behavior can be
identified. We report in Fig.~\ref{fig.results}(B,2) the fluctuations
expected for the temperature $T=84$~nK (obtained from the fit of the
momentum distribution with QMC). They are in agreement with the
measured fluctuations. Note however that, for these parameters, the
uncertainty of thermometry based on fluctuations is about 30\% (see
gray lines in Fig.~\ref{fig.results}(B,3)). The density profile
expected for $T=84$~nK is not far from the measured one (see
Fig.~\ref{fig.results}(B,4)).  In contrast to case (A), Yang-Yang
thermometry based on the profile is less sensitive to tiny
modifications of the wings, since the profile of the high-density
regions is also affected by the temperature. Thus, a thermometry based
on the profile is expected to be quite precise and the potential
roughness is expected to have a smaller effect.  A Yang-Yang fit to
the experimental profile yields a temperature of $76$~nK, different
only by 10\% from the temperature deduced from the momentum
distribution.

\section{Conclusions and prospectives}
\label{sec.conclusions}

We have shown that for $t<10^6$, a simple 1D classical field
approximation where the momentum distribution 
is computed without energy cutoff,
fails to describe the momentum
distribution of weakly interacting 1D Bose gases in the crossover from
ideal Bose gas to quasi-condensate. Thus, we performed QMC
calculations to investigate the crossover.  Experimentally, we measure
the momentum distribution with the focusing technique, which is
improved by a guiding scheme for enhanced resolution.  We show that
the temperature deduced from the momentum distribution is in agreement
with an independent fluctuation-based thermometry.  We find that the
result of Bogoliubov theory is not appropriate for parameters that are
close to the ones relevant to this article, even when the cloud lies
quite deeply in the quasi-condensate regime.  In~\cite{Davis2012}, it
has been proposed to use the mean kinetic energy deduced from measured
momentum distribution in order to extract the temperature using the
modified Yang-Yang calculation.  For the data presented in this
article, with a SNR of about 50, the measured momentum distribution
shows tails compatible with a $1/p^2$ behavior both in the
quasi-condensate regime and in the degenerate ideal Bose gas
regime. Therefore, extraction of the mean kinetic energy from the data
is impossible.  We argue that this technique was applicable in
~\cite{Davis2012} because of the more 3D nature of the gas: for quasi
1D gases, as discussed in section~\ref{sec.resultsB}, the contribution
of the transverse excited states leads to fast decaying tails, which
enable the extraction of the mean kinetic energy per particle.  We
also remark that neither our measurements nor the QMC calculations
display the asymptotic behaviour of $1/p^4$, which only appears at
even larger momenta for our parameters.  On the other hand, we believe
that this behaviour could be more readily discernable in the strongly
interacting regime, where the $1/p^4$ tail should contain a larger
proportion of the atoms.

This work opens many perspectives for the study of 1D gases.  First of
all, we show that momentum distribution measurements provide a precise
thermometry for 1D Bose gases. While in this article we use QMC
calculations to fit the temperature of experimental data, a rough
estimate of the cloud temperature could be performed with a lower
numerical cost.  Indeed, the comparison with the exact QMC momentum
distribution shows that a model which combines the ideal Bose gas
theory and the classical field approximation gives the correct width
within 20\% precision.  Second, momentum distribution measurements are
essential to characterize more complex systems.  For instance, in the
presence of a lattice, it enables the investigation of the correlation
properties at the Mott and/or pinning transitions.  Contrary to Bragg
spectroscopy \cite{Richard2003, Clement}, where a single momentum
component is probed at each shot, the focusing method as well as time
of flight method gives access to the whole distribution at the same
time, which allows for noise correlation measurements in momentum
space~\cite{correlk,Folling2005}.  Finally, recording the time
evolution of the momentum distribution is essential to monitor the
non-equilibrium dynamics and address the issue of thermalization in 1D
closed quantum systems. For example, the dynamics resulting from a
quench of the 1D coupling constant can be investigated.  The measure
of the momentum distribution of impurities would also permit the study
of impurity dynamics, which is currently under intense
investigation~\cite{catani}.

\begin{acknowledgments}
The authors thank Karen Kheruntsyan for providing the data from the
Yang-Yang calculations.  This work was supported financially by the
Triangle de la Physique, by the ANR grant ANR-09-NANO-039-04, the
Austro-French FWR-ANR Project I607, the ARC Discovery Project Grant
No. DP110101047, and the CoQuS Graduate school of the FWF.

\end{acknowledgments}


\begin{thebibliography}{10}
\bibitem{TQ}
Quantum fluctuations lead to an algebraic decay of $g^{(1)}$ as
  $(z/\xi)^{-\sqrt{\gamma}/2\pi}$ for distances $\xi<z<\hbar^2/(mk_BT\xi)$,
  where $\xi = \hbar/\sqrt{m\rho g}$ is the healing length. For the decay to be
  noticeable over this range, one requires $k_B T \propto \rho g
  \exp(-2\pi/\sqrt{\gamma})$.
\bibitem{SP}
In terms of the analog quantum problem, $g^{(1)}(z)=\Sigma_j A_j
  e^{-(\epsilon_j-\epsilon_0)|z|}$, where the sum is carried out over the
  eigenstates of energy $\epsilon_j$, and the factors $A_j$ are obtained by
  wavefunction overlaps. The Fourier transform is a sum of Lorentzian, which
  shows $1/p^2$ tails (uniform convergence is ensured by the fast decay of the
  $A_j$'s at large energy).
\bibitem{ntof}
This relation is easily established using the Wigner function representation,
  since its evolution under a harmonic potential and its free evolution are
  those of a classical distribution.
\bibitem{anharm}
Deviations from the harmonic potential by about 30~nK is sufficient to explain
  the presence of those wings. This value is consistent with the residual
  roughness we estimated from the measured density variations of very elongated
  quasi-condensates confined in a longitudinal quartic potential.
\end{thebibliography}

\end{document}